# Economic Policy Challenges for the Age of AI[1]

Anton Korinek (University of Virginia, CSH Vienna, and GovAI)

September 2024

This paper examines the profound challenges that transformative advances in AI towards Artificial General Intelligence (AGI) will pose for economists and economic policymakers. I examine how the Age of AI will revolutionize the basic structure of our economies by diminishing the role of labor, leading to unprecedented productivity gains but raising concerns about job disruption, income distribution, and the value of education and human capital. I explore what roles may remain for labor post-AGI, and which production factors will grow in importance. The paper then identifies eight key challenges for economic policy in the Age of AI: (1) inequality and income distribution, (2) education and skill development, (3) social and political stability, (4) macroeconomic policy, (5) antitrust and market regulation, (6) intellectual property, (7) environmental implications, and (8) global AI governance. It concludes by emphasizing how economists can contribute to a better understanding of these challenges.

## I. Introduction

The rapid rise of artificial intelligence (AI) will present significant challenges for economic policy. The release of large language models such as ChatGPT in late 2022 has heightened public awareness of AI's potential and stimulated discourse on its societal impacts, yet it represents just one steppingstone on the larger trajectory of AI advancement. As AI capabilities continue to grow, it becomes increasingly crucial to re-evaluate our economic and policy frameworks to ensure their relevance and appropriateness for the Age of AI.

This paper examines the economic implications of advanced AI, with particular emphasis on the potentially impending development of Artificial General Intelligence (AGI) – AI systems capable of performing intellectual tasks at or above the human level across domains. I argue that the advent of AGI could fundamentally alter the dynamics of our economic system, necessitating a comprehensive reassessment of economic theory and policy.

[1] This paper was prepared for a conference organized by the Peterson Institute for International Economics and the International Monetary Fund on "Rethinking Economic Policy: Steering Structural Change." The author would like to thank Justin Bullock, Duncan Cass-Beggs, Tom Cunningham, Bill Gates, Tim Laseter, Sam Manning, Donghyun Suh, Joe Stiglitz, and Petia Topalova as well as the participants of the conference for their thoughtful comments and discussions. Financial support from the University of Virginia's Bankard Fund for Political Economy and Darden School of Business is gratefully acknowledged.

I start with a brief description of the trajectory of technological advances driving AI progress, emphasizing the exponential growth in computational power and increases in algorithmic efficiency. Then I draw on the perspectives of eminent AI researchers and industry leaders to lay out potential trajectories for AGI development and discuss key indicators that may signal its approach.

My main focus is the economic implications. I argue that the emergence of AGI will mark the end of the Industrial Age and usher in a new economic era, just like the Industrial Age ended the Malthusian era. This change in paradigm will involve a significant shift in what factors of production are relevant for the economy, away from labor to reproducible factors such as computational resources and robots, with consequent implications for productivity and output growth.

The paper proceeds to raise eight challenges that advanced AI poses for economics and economic policy. First, I discuss the challenge of income distribution and inequality in a world where human labor may become increasingly substitutable. Second, I assess the devaluation of human capital and the need to recalibrate approaches to education and skill development. Third, I focus on the resulting challenge to social and political stability. Fourth, I discuss macroeconomic considerations, including shifts in aggregate demand in a highly automated economy in which labor's relevance diminishes, and potential adaptations for monetary and fiscal policies.  Fifth and sixth, I bring up the implications for antitrust policy and intellectual property regimes. Seventh, I examine the environmental challenges from rapid AI development. Eighth, I raise international economic policy and governance challenges, including the management of AI-driven geopolitical dynamics and the mitigation of potential global disparities in AI capabilities. Successfully preparing our economy and society for the potential of AGI will require meeting all the described challenges – and likely many more.

The paper concludes by reflecting on the role of economic analysis in addressing these challenges, and by emphasizing how AI itself can be leveraged as a tool for economic analysis. Throughout the paper, I emphasize the importance of proactive preparations for the challenges and opportunities presented by advanced AI. As we approach a potentially transformative technological shift, it is imperative that we examine and adapt our economic policies to the novel challenges of the AI age. This paper aims to contribute to this crucial research agenda.

## II.   Technological Capabilities

The field of AI has witnessed unprecedented growth in recent years, with progress accelerating at a pace that challenges our traditional understanding of technological advancement. This section explores the key drivers behind this rapid progress and examines the potential trajectories for future AI development.

### 1.  Exponential progress in computing

Moore's Law has long been a key benchmark for progress in the domain of computation. Translated into economic terms, Moore (1965) predicted a doubling in the efficiency of microprocessors every two years. This regularity has approximately held for more than a half-century now, enabling rapid efficiency gains in computing, which in turn underpinned the rise of AI in recent decades.

The pace of growth in the complexity of AI systems, however, has far outstripped even these rapid efficiency gains. As observed by Sevilla and Roldan (2024), the amount of computational power (or "compute") employed in training cutting-edge AI systems has doubled every six months over the past decade. This trend is visually represented in Figure 1.

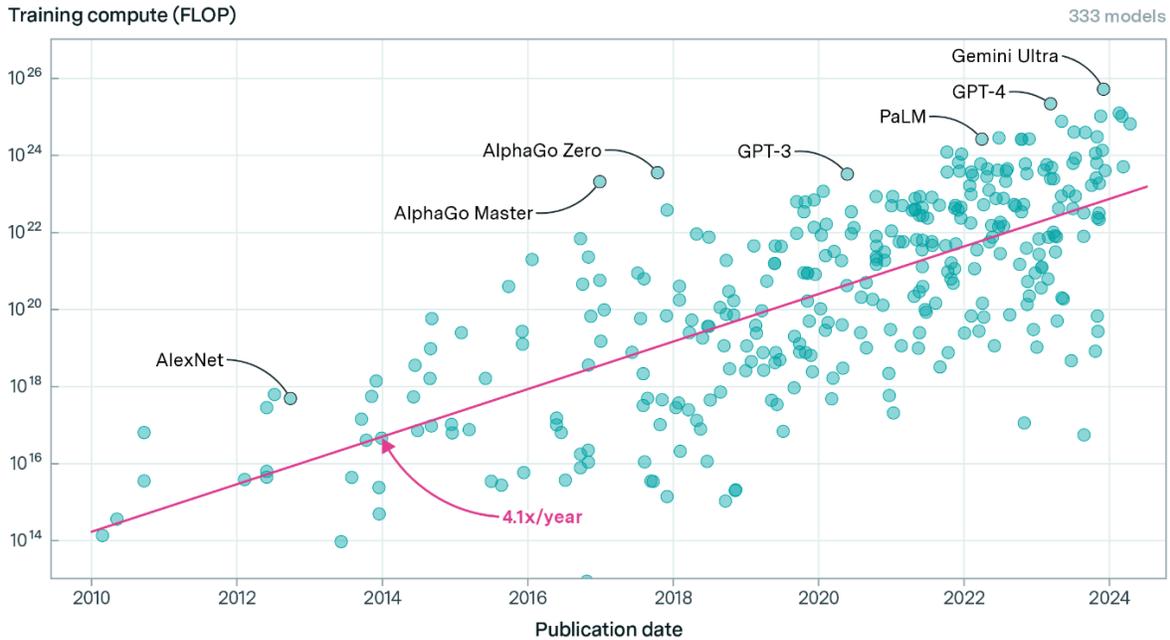

Figure 1: The training compute employed by the most cutting-edge AI models doubled on average every six months for the past 15 years. Distributed under a CC-BY 4.0 license by Epoch.

The training compute of the systems depicted in the figure has grown much faster than what is implied by Moore's Law because the investments in compute by frontier AI companies have on average tripled every year over the period.[2] Estimates of the cost of current frontier AI systems are in the range of hundreds of millions of dollars, potentially soon surpassing the billion dollar threshold. Industry observers predict that the exponential growth of the compute invested in frontier AI models will continue for at least another three to five years, and possibly longer, given the recent economic success of AI models (Sevilla et al, 2024). If current trends continue, we may see trillion dollar training runs for frontier AI models by the end of the decade.

The rapid growth in the computational resources deployed is complemented by significant improvements in algorithmic efficiency. Ho et al (2024) report that the algorithmic efficiency with which AI systems utilize compute has grown at a rate of about 2.5x per year. When multiplying these efficiency gains with the 4x increase in compute use stemming from Moore's Law and increased investment in compute, these factors result in a staggering 10x increase in the effective compute used for training frontier AI systems every single year. In addition, the most recent generation of AI models that can perform reasoning-related tasks, such as OpenAI's o1, also employ substantial amounts of compute at the inference stage, i.e., when they responding to user inquiries.

---

[2] A doubling of compute every six months implies a 16-fold increase every 2 years. Since Moore's Law implies "only" a doubling of chip efficiency every two years, an 8-fold increase in spending on compute is required every two years to achieve the observed overall growth of compute – almost amounting to a tripling per year.

The implications of this rapid growth in compute are profound. As AI systems become more powerful, they can tackle increasingly complex tasks, leading to breakthroughs in more and more domains of human cognitive work. The complexity of frontier AI models is rapidly approaching our best estimates of the complexity of the neural networks in the human brain (Carlsmith, 2020). Moreover, advances in robotics such as humanoid robots are also proceeding at a rapid pace (see, e.g., Du et al., 2024).

## 2. Compute and capabilities

While inputs to AI and algorithmic efficiency are important, what matters from an economic perspective is how these inputs translate into outputs, i.e., into AI systems' capabilities. This is what we turn to next.

**Scaling laws**  Advances in AI capabilities follow predictable regularities – so-called scaling laws – that describe how much better AI systems become at prediction as the number of model parameters and the amount of training data increase (Kaplan et al, 2020; Hoffman et al, 2022). Together, the parameter count and amount of training data determine how much compute is needed to train a model. The described scaling laws have held for more than a decade now and are at the heart of the strategies employed by frontier AI labs in their pursuit of AGI. The predictive power of these laws allows researchers to estimate how much improvement in AI capabilities can be expected from further increases in computational resources, guiding both research and investment by AI labs. More recently, OpenAI (2024) found that scaling laws apply not only to the model training stage, but also when letting AI models deliberate before giving answers at the inference stage, implying that more compute is predictably associated with better answers.

More generally, neural networks, the fundamental building blocks of modern AI systems, are universal approximators of arbitrary functions (Cybenko, 1989; Hornik et al, 1989). This means that, in theory, they can perform any information processing task arbitrarily well, given sufficient resources and training data. The experience of the past decade has shown that they are in fact very efficient function approximators for a wide range of real-world use cases. This property underpins the potential of AI systems across a wide range of applications – including the possibility of performing all the intellectual functions of the human brain.

## 3. Predictions of AGI

One of the crucial questions regarding the future impact of AI is whether and when AI systems will match human levels of intelligence across all cognitive tasks that humans can perform, as this would create the potential for widespread labor substitution. AI systems that can do this are termed artificial general intelligence (AGI) to distinguish them from narrow AI systems that can only perform individual tasks or a small range of tasks, such as image recognition or voice transcription (Morris et al., 2024).

Just a few years ago, all AI systems were narrow. However, modern large language models (LLMs) and other types of generative AI since the early 2020s are increasingly general purpose, performing a growing range of cognitive functions, from creative tasks to simple forms of reasoning.

The rapid progress over the past few years has compressed the timelines for achieving AGI. In the 2010s, the median estimate of AI researchers for when AGI would be reached was in the second half of the 21st century (Grace et al., 2018). More recently, a growing number of AI researchers and industry leaders have offered much shorter timelines. Geoffrey Hinton, one of the three "godfathers of deep learning" who won the Turing award for their contributions to the field, proclaimed that "[he had] suddenly switched [his] views on whether these things are going to be more intelligent than us" and expects AGI to be reached in "5 to 20 years but without much confidence [since] we live in very uncertain times" (Hinton, 2023). Likewise, Sam Altman, CEO of OpenAI, stated that "AGI will be a reality in 5 years, give or take" in early 2024. These predictions are also in line with the expectations of the general public (Pauketat et al., 2023).

AGI would be transformative for our economy and society, creating both opportunities and risks of unprecedented scope. This has led to a growing sense of alarm among technology experts who see potentially radical changes occurring within a short period. They see the exponential growth of AI and associated scaling laws and wonder why the world is not more alarmed.

Geoffrey Hinton's warnings are akin to Albert Einstein's letter to US President Franklin Delano Roosevelt, in which he warned of the possibility that nuclear power could be harnessed for atomic weapons, giving rise to the nuclear age. There was significant uncertainty about how realistic the predictions of Einstein and other experts were at the time, and how soon their predictions would be realized. In 1933, Ernest Rutherford called the notion that nuclear power could ever be harnessed "moonshine." We are facing similar uncertainty today, but the scaling of AI is nonetheless proceeding relentlessly.

## 4. Harbingers of AGI

Economists have an important role to play in shaping the direction of progress towards AGI (Korinek and Stiglitz, 2022) and in helping our society to prepare for and adapt to the resulting disruptions as well as to integrate the increasingly capable AI systems into our economic system.

Given the uncertainty about whether and how quickly the development of AGI might come to pass, it is important to closely follow advances in AI and regularly update predictions. To do so, it is useful to monitor four key categories of indicators:

1. **AI Research Breakthroughs**: Tracking major advances in AI research and development is perhaps the most direct indicator of progress in AI, although it is difficult to appreciate the overall impact of a series of relatively small individual advances. For example, leading AGI labs are currently focused on improving the reasoning capabilities of AI systems – a weak point of LLMs. This may require approaches that go beyond the current architecture of LLMs. Some of the recent advances in this domain were Google DeepMind's AlphaProof and AlphaGeometry 2, which solved math problems of equivalent difficulty to Silver Medalists at the 2024 International Math Olympiad (DeepMind AlphaProof and AlphaGeometry Teams, 2024). Moreover, monitoring the trajectory and speed of AI models improvement allows observers to recognize trends and make better-informed predictions.

A particularly important type of research breakthrough are advances that automate the research process itself, which can significantly accelerate the creation of better AI systems in the future. If AI systems can produce better AI systems with less and less human input, it marks an important step towards AGI. AI systems are already contributing to their own development. Peng et al. (2023) demonstrate that access to AI-powered coding assistants allows programmers to complete coding tasks 56 percent faster – and this also applies to the programmers of future AI systems. Lu et al. (2024) describe an AI agent that they term "AI scientist," which can autonomously produce ideas for papers in computer science, execute them, and write up the results, giving rise to novel research insights. AI is also being used in the design of AI chips, potentially accelerating hardware development for AI systems (Mok, 2024).

2. **AI Product Releases**: Tracking the performance of new AI products on established benchmarks can provide additional concrete evidence of AI capabilities. This is highly useful for everyone who performs cognitive work – for example, I myself continually track progress in generative AI systems to better understand how I can optimally deploy them in my research. This allows me to have a first-person perspective on how AI is revolutionizing the conduct of economic research (Korinek, 2023a, 2024). There is no good substitute for having this type of subjective experience.

3. **AI Deployment Across the Economy**: Lastly, monitoring the adoption and impact of AI technologies in work settings can offer a real-world perspective on the practical AI advancements. This can most easily be observed in one's own area of work – for example, in economics, generative AI has been adopted heavily by younger researchers such as PhD students and assistant professors, whereas older generations of economists lag behind, making it harder for them to assess advances in AI. Oftentimes I learn about new AI capabilities by looking over the shoulders of my younger colleagues. Naturally, a complete picture can only be obtained by monitoring the deployment of AI across different sectors of the economy to get a better understanding of what tasks can or cannot be performed by AI and how close we are to AGI (see, e.g., Bonney et al, 2024).

4. **Financial Market Trends**: Financial markets are a mirror of the expectations of their participants and therefore reflect public expectations of AGI. In primary markets, the expectation of AGI is likely to lead to a growing influx of investments into AI, both on the hardware and software side. In secondary markets, stock prices of existing corporations that produce, incorporate, or otherwise benefit from AI are likely to see large run-ups in asset prices. Moreover, interest rates may rise in parallel with a pick-up in growth expectations (Chow et al., 2023). Although financial markets are inherently noisy and prone to booms and busts, their forward-looking nature provides a useful complement to real indicators.

## III. Economic implications of AGI

This section discusses the implications of AGI for economics and economic policy. For the sake of clarity, I will cut through the uncertainty regarding the development of AGI and take it as given that AGI will be reached within the foreseeable future, in line with Hinton's prediction. Moreover, I will assume that progress in robotics will proceed, and that AGI-powered robots can perform any physical work that humans can perform soon after AGI is reached.[3]

Rigorously preparing economic policy for the Age of AGI will require scenario planning to examine the impact of alternative economic policies under different scenarios for technological advancement. Korinek (2023b) and Korinek and Suh (2024) provide examples. However, even if full AGI is not achieved in the near future, the economic challenges discussed below are directionally correct and relevant.

## 1. Historical economic paradigms

To understand the economic implications of a change as transformative as AGI, it is useful to take a step back and look at the longer-term trajectory of our economic system, starting with the Malthusian Age. I will then explore what changed in the Industrial Age and what is likely to change in the Age of AGI, focusing on the effects on output and wages.

**The Malthusian Age**

In the Malthusian Age, the economy was characterized by a production function of the form

$$Y = A \, F(T, L)$$

where technology, A, was largely stagnant (or growing very slowly), land, T, was the scarce resource, and humans who provided labor, L, were reproducible. In this era, population growth was constrained by the resources the economy could produce. Given the fixed factor land, an increase in population implied a decreasing marginal product to human labor. Humans reproduced to the point where our marginal product equaled our subsistence level. A larger population could not be supported so any additional population growth led to destitution, hunger, and death. Given this situation, land was the most valuable production factor – far more valuable in relative terms than it is today. Humans earned a subsistence income that was just enough to survive. From a purely economic perspective, the marginal human was dispensable.

**The Industrial Age**

The economy of the industrial age has been the main focus of modern economics – it captures the world we are still living in. It is characterized by a production function of the form

$$Y = A \, F(K, L)$$

where growth in technology, A, is the key driver of economic growth, capital, K, is the reproducible factor and accompanied the growth in A, and labor, L, grows so slowly that is usually taken as exogenous. Since the Industrial Revolution, output, Y, in advanced countries has grown about 20-fold, driven by comparatively rapid technological advances.

---

[3] Robotics has recently experienced breakthroughs by combining robots with more capable brains in the form of modern AI systems. Robots that can perform any human tasks will likely take longer than 5 years, i.e., they will materialize more slowly if AGI happens soon. However, AGI will likely design robots that can perform any physical task if humans have not gotten their first.

Given these advances, labor suddenly became the scarce factor of production and experienced large increases in its marginal product in parallel with the 20-fold increase in output, giving rise to the level of wages that we see today. In short, labor transitioned from being dispensable to becoming the bottleneck production factor of the economy.

**The Age of AGI**

When AGI is reached, AI systems can, by definition, perform any cognitive work that humans can perform, and, per our earlier assumption, will soon perform any physical work as well, giving rise to a production function of the form

$$Y = A\, F(K, L + M)$$

in which the new variable, M, captures machines, in the form of AI compute and robots. In this simple model, machines are a perfect substitute for human labor, i.e., for both human compute and human physical capabilities.[4]

In an AGI-powered world, growth in technology A will also accelerate, as artificial brainpower and robots can drive scientific progress and innovation. Both traditional physical capital K and the new machines will be reproducible resources. The distinction between the two will increasingly blur. They will accumulate without bounds and generate ever more economic capacity, in the spirit of Romer (1986)-style AK-models.

An immediate implication is that the growth rate of output will rise. If technology is accelerating and all factors of production are reproducible, there are no more bottlenecks to output growth in the form of fixed factors.

The implications for the marginal product of labor are also straightforward. In the Malthusian age, land was the bottleneck factor, and the marginal product of labor was competed down to the human subsistence level by the cruel process of humans multiplying until the available living space and arable land were exhausted. The Industrial Age ushered in a Golden Age for humans: labor was suddenly the bottleneck factor, and the average returns to labor grew about 20-fold above our subsistence needs. In the Age of AGI, there's no reason for labor to continue to play such a special role. If AI can solve all computational problems that our brains can solve, why should human labor have a special status in our economic system? Labor would be a perfect substitute for machines so the two would be equally scarce. As more and more machines are accumulated, the value of labor and by extension wages will diminish. This will fundamentally challenge our present system of income distribution.

However, the divergence of output growth and the marginal product of labor also suggests the contours of a solution. What is needed to create shared prosperity is to take a small sliver of the rapidly rising output from AGI and give it to the workers whose labor market value is diminished. This creates the potential for a Pareto improvement, although such a solution may face significant political economy challenges (see, e.g., Bell and Korinek, 2023).

---

[4] A more comprehensive production function under AGI could distinguish labor into its components physical and cognitive labor and observe that AI and robots respectively substitute for each of the two components. See, e.g., Growiec et al (2024).

## 2. AGI scenarios

To illustrate how the forces described above unfold in a comprehensive macroeconomic model, I provide an overview of the four main examples from "Scenarios for the Transition to AGI" by Korinek and Suh (2024). Our analysis is grounded in a task-based framework in the spirit of Zeira (1999) and Acemoglu and Restrepo (2018). Our main innovation is to represent work as consisting of tasks varying in computational complexity that can be automated only when AI reaches a sufficient level of capabilities. This approach allows us to model how exponential advances in AI progressively automate more tasks and, if there is a maximum level of task complexity and AI surpasses it, to capture the advent of AGI.

We consider four distinct scenarios, each based on different assumptions about the distribution of task complexity and the rate of technological progress. In all scenarios, we assume that an automation index, representing the maximum complexity of automatable tasks, grows over time at an exogenous rate, reflecting reflects the observed trends in computing power and AI capabilities discussed before. The key differentiator between scenarios is the distribution of task complexity.

1. **Business-As-Usual Scenario:** This scenario assumes that task complexity follows a Pareto distribution with an infinite right tail, and the fraction of non-automated tasks declines at a constant rate. This implies that there are always tasks that only humans can perform or that we choose not to automate. Under our assumptions, output grows steadily at about 2 percent per year. For our parameterization, both the returns to capital and the wage bill rise indefinitely, approximately in tandem.
More generally, there is a race between automation and capital accumulation – automation tends to pull down wages whereas the accumulation of capital that is complementary to the remaining tasks for labor increases wages. If automation proceeds too quickly, wages may decline, even if full automation is never reached.

2. **Baseline AGI Scenario:** In this scenario, the task complexity of human work follows a distribution with a finite upper bound, and full automation is reached within 20 years, in line with Hinton's upper range for when AGI will be reached. The economic implications are stark: wages grow during the initial period but collapse before full automation is achieved. After the wage collapse, labor and compute earn equal returns. The economy transitions to a steady-state growth of 18 percent per year, but the labor share declines dramatically as automation progresses.

3. **Aggressive AGI Scenario:** This scenario is similar to the Baseline AGI scenario but assumes a shorter-tailed distribution, with full automation reached within 5 years, in line with Hinton's shorter estimate for AGI. The wage collapse starts much earlier, in year 3. The economy experiences a more rapid transition to high growth rates, and labor becomes perfectly substitutable with compute very quickly. This scenario results in a quick concentration of returns to capital owners.

4. **Bout of Automation Scenario:** The final scenario assumes that the economy experiences a large bout of automation in the short run, for example, because cognitive labor is automated, but that there remains an infinite tail of tasks that either cannot be automated or that we choose not to automate. Initially, rapid automation leads to a wage collapse similar to the Aggressive AGI scenario. However, as the economy accumulates more capital, labor becomes scarcer again. Eventually, wages rise above the return to capital and start growing again. This scenario illustrates the possibility of a temporary labor demand collapse followed by recovery.

These scenarios highlight the critical role of (i) the complexity required to automate different work tasks and (ii) the speed of progress in AI in determining the dynamics of output and wages during the transition to AGI and beyond. They underscore the importance of understanding the nature of human cognitive capabilities and any potential limits of machine intelligence in shaping our economic future. The stark differences between the scenarios also emphasize the importance for economic policy to be adaptive as we navigate the uncertain path towards AGI.

Some argue that we need not be concerned about the fate of labor since the comparative advantages of humans and machines will always allow for gainful trade between the two (see, e.g., Smith, 2024). Although it is true that agents with different capabilities – like humans and robots – will have comparative advantages, this does not prevent wage declines from technological changes. Comparative advantage just tells us that two agents with different endowments or capabilities benefit from trade – it does not tell us how these benefits will evolve over time. Under AGI, the relative *terms-of-trade* of humans may deteriorate significantly and may imply significant income losses for the average worker. As an analogy, horses and humans have always had comparative advantages, and they still do, yet the combustion engine has greatly undermined the economic value of horses.

## 3. Scarce factors or post-scarcity society?

An influential line of thought in Silicon Valley, fostered by science fiction works such as the "Culture" series or "Star Trek," is that AGI may usher in a post-scarcity world. Post-scarcity is usually defined as a hypothetical economic condition where any goods required to meet human needs can be produced in great abundance and become freely available to all. Some argue that in a post-scarcity world, traditional economic concepts like supply and demand, scarcity-based pricing, or money might become obsolete.

In economics, a good is typically considered scarce whenever its price is greater than zero. This use of the term "scarce" differs from the use in common language, where it often simply refers to the opposite of "abundant." In economics, a good can be scarce (i.e. have a positive price) and relatively abundant at the same time, resulting in a price that is low but still positive. The economic notion of scarcity is useful because a positive price captures that there are trade-offs in the allocation of resources. Despite the great abundance of material goods in our lives today compared to Malthusian times, resources are still scarce, and we continually need to decide how to allocate them – for example, whether to spend a hundred million dollars on training the next AI system or on relieving hunger in Africa.

From my perspective, scarcity in the economic sense will always remain, and it is important to understand its implications in the Age of AGI. From an economic perspective, our best hope is that AGI will usher in a world in which many of the bottlenecks to growth are overcome and the economy will grow faster – compute and robots will produce more compute and robots at an increasing pace. As a result, output will be much larger bringing greater abundance (Korinek and Trammell, 2023). Compute and robots may become cheaper compared to resources that are in more limited supply, but they will still be scarce in the sense that their price will be greater than zero. This implies that their distribution and use will still be subject to trade-offs, and these trade-offs will play a central role in economic policy.

Speaking of prices, a parallel line of thought in Silicon Valley is that the prices of all goods will decline dramatically in an AGI-powered world (see, e.g., Altman, 2021). However, what matters for economic allocations is relative not absolute prices. If the relative price of one good in terms of a second good declines, the relative price of the second good in terms of the first good necessarily rises. If AGI materializes, one possibility is that the *relative* prices of computers, robots, and human labor (i.e., wages) decline whereas the *relative* prices of housing, energy, and food may rise.

To systematically analyze which prices will rise in relative terms under AGI, it is useful to distinguish between the factors of production that will be reproducible and those that will remain irreproducible and thus bottlenecks for the expansion of output. During the Industrial Age, capital was reproducible, but labor was not. As a result, wages along with the relative price of labor-intensive goods and services continuously rose. In recent decades, as technology automated unskilled work, skilled labor and its output rose in value. As a first approximation, AGI plus advanced robotics may make all labor reproducible, reducing its value.

A more fine-grained approach requires that we disentangle which specific parts of capital are irreproducible, and which types of human labor will still be in demand. We focus on disentangling what has traditionally been lumped together as capital in the remainder of this subsection and analyze labor post-AGI in the ensuing subsection.

Many forms of capital will be reproducible, despite their currently high value and/or production bottlenecks. For example, an Nvidia Blackwell GB200 chip costs around $60,000 at the time of writing, in part because of the limited capacity to produce such cutting-edge chips. However, I am certain that the same amount of computing power will be available at a significantly lower cost in a few years. The same holds for robotics. These forms of reproducible capital may initially be expensive but will become cheaper over time.

Irreproducible factors are different. Land is one of the most fundamental irreproducible factors traditionally lumped into capital. Physical space on earth is inherently limited and cannot be reproduced so it is likely to go up in value as the economy grows and other factors reproduce. However, high land prices have traditionally given rise to land-saving innovations (such as skyscrapers) and may eventually induce machines to settle beyond earth. Other physical factors that are in limited supply include energy and raw materials (for example, precious metals and rare earths). Over time, we will become increasingly better at accessing the two and at transforming them into each other as needed, for example, by unlocking energy production from nuclear fusion. Eventually, matter or equivalently energy may become the final irreproducible factor in the universe.[5]

There will also be traditional goods that are inherently in limited supply, for example, original pieces of fine art like the Mona Lisa, or historical artifacts. Even though they can be copied almost perfectly, humans still care about the attribute of something being original, in part because this confers scarcity and is by extension a status symbol.

---

[5] Einstein taught us that $E = mc^2$, i.e., that there is an equivalence between energy and mass, exploited for example in the production of nuclear power.

Finally, our legal system creates artificial scarcity through intellectual property rights, which impose excludability on inherently non-rivalrous goods. In economics, non-rivalry refers to goods whose consumption by one individual does not reduce their availability to others, such as ideas, information, and digital content. However, copyright laws, patents, and other forms of intellectual property protection make these naturally abundant resources artificially scarce, allowing creators and inventors to control access and monetize their work. While this system provides economic incentives for innovation, it creates artificial scarcity that leads to inefficiencies. For example, life-saving medical treatments protected by patents may be priced out of reach for many, despite negligible production costs. As AGI accelerates the pace of innovation and idea generation, we may find it useful to reconsider the balance between incentives for innovation and the widespread societal benefits from freely accessible knowledge and technology (Shapiro 2008).

## 4. Labor post AGI

In a world where AGI and advanced robotics can perform any cognitive and physical task that humans can, labor will lose its special status of being the main irreproducible factor in the economy. The value and role of labor will diminish in such a world.

However, several factors will sustain some demand for human involvement in the economy, at least in the short to medium term. In the following list, the first four factors can be classified as *temporary technical and social barriers*, which reflect temporary limitations and factors that may diminish over time as AI technology advances and society adapts. The subsequent four factors can be classified as *fundamental human-centric aspects* of labor, which are more deeply rooted in human nature, values, and societal structures and may rely on a valuable human element for a longer period.

1. **Production and diffusion lags:** Once machines reach human levels of performance and become competitive in cost, there will be significant economic incentives to deploy them throughout the economy. However, the production, distribution, and adoption of large numbers of AI systems and robots throughout the economy will take time, implying that human labor will be employed to perform tasks that can in theory already be done by machines. Part of the reason for such lags is that many consumers and producers as well as employers and employees have long-lasting relationships that involve important social dimensions. However, as AI systems and robots become cheaper and are produced on a greater scale, it will become less and less economical to employ humans when machines can perform the same jobs better and more efficiently. The stronger the competitive forces in a sector, the faster machines will be adopted. For example, consultants or workers in car manufacturing may see their jobs replaced faster than government workers.

2. **Implicit knowledge:** AGI together with human-level robots do not imply that machines can immediately perform all human jobs. Even though such machines will be able to rapidly internalize any knowledge that has been written down in explicit format, they will require training and practice to acquire the implicit knowledge necessary for many jobs. In sectors with low stakes and significant volume, training and practice will happen fast. For example, AI-powered robots could quickly be trained as super-human cleaners once technically feasible and could be widely deployed once economically viable. On the other hand, it might take longer for robots to acquire human-level performance in smaller markets, say, the restoration of antique clocks (although they may figure it out quickly with super-human cognition). However, even in such niche jobs, it is likely that transferring the implicit knowledge to a machine will be quicker – and more scalable – than transferring it to another human.

3. **Trust and perceived human superiority:** In many sectors, a lack of trust in machine capabilities and perceived human superiority will slow the adoption of machine services and preserve a role for workers. There will be wide heterogeneity in attitudes – from innovators and early adopters to laggards (as described by Rogers, 1962). In many sectors, professionals will increasingly act as a "human veneer" for AI – for example, doctors, lawyers, consultants, or financial advisors will simply convey AI diagnoses and advice to their clients, knowing that adding their personal judgment will often make things worse. Ensuring a sufficient level of cybersecurity will be an important element of creating trust. Over time, trust in AI will grow, and human jobs will progressively be phased out. In part, this will be a generational transition, with younger consumers being more comfortable receiving AI services.

4. **Laws and regulatory reasons** A significant number of jobs are protected from AGI encroachment by laws and regulations, even though AI systems may perform the job in a superior manner. According to current law, politicians, lawmakers, judges, or company officers, all need to be natural persons. Moreover, regulations require many medical and legal services to be performed by human doctors and lawyers, even if it is just as a veneer for AI decision-making, as noted above. Eventually, society may rethink which positions require natural persons and update laws and regulations – for example, a growing number of jurisdictions may allow AI company officers to increase efficiency or AI doctors to improve access to healthcare.

5. **Authenticity of human connection:** In many services consumers may demand human providers because they perceive humans as more authentic than machines that render the exact same service, for example as humanoid robots that simulate their emotional attunement. This may be the case even if the machine service is objectively indistinguishable or superior. For example, consumers may prefer human therapists knowing that only humans can authentically share their emotional pains. They may prefer to send their children to human childcare providers because of the (hopefully) authentic human connection that they offer. As AI improves, consumers will have to forgo increasingly super-human service quality to satisfy their desire for authentic human connection, and demand for human services of this category may decline.

6. **Human Identity:** A related category of demand for human labor post-AGI will arise from competitive or performative jobs in sports and the arts, in which the human identity of the performer matters. Soccer fans will not be equally impressed if superhuman robots beat a human team (although new robot soccer leagues may also be set up), and culture fans may not be impressed if a robot outperforms a prima ballerina. In these types of activities, part of what matters is that spectators can see a fellow human with whom they can identify and who performs activities that push human limits.

7. **Religion:** Religious beliefs require some human services, for example, religious acts, to be performed by humans since adherents of many religions believe that humans distinguish themselves from all other entities, for example, by having a soul. This would be another source of demand for human services. Not all religious believers will be convinced by Alan Turing (1950)'s suggestion that humans developing intelligent machines are creating potential vessels for souls in the same manner as humans who are pro-creating do.

8. **AI alignment:** Even in a world where AGI can perform all tasks better than humans, AI alignment – ensuring AI systems behave in accordance with human values and intentions – may require a human perspective. Only humans can serve as the ultimate arbiters of what constitutes "aligned" behavior as societal values continually evolve. This role leverages not intelligence, but preferences based on humans' lived experience. Relying on AI for alignment could create a closed-loop system that might drift away from human values over time. Human oversight may thus be needed as an external anchor, ensuring that the alignment process itself remains aligned with human interests.

While these factors may sustain some demand for human labor in a post-AGI world, many of them represent transitory phenomena rather than permanent safeguards for human employment. Over time, continued advances in AI capabilities will make inroads and may eventually erode these niches for human labor. As AI systems become increasingly sophisticated, they will more efficiently acquire implicit knowledge, gain societal trust, and provide authentic-seeming interactions. Simultaneously, human adaptation to AI services and broader societal changes will diminish the perceived need for human involvement in many areas. Legal and regulatory frameworks will evolve to accommodate the new realities of an AGI-driven economy. Even in areas like religion and AI alignment, the unique value of human input may diminish as AI systems become better at understanding and replicating human values and experiences. Consequently, while the transition may be gradual, the long-term trajectory points towards a significant decline in the role of human labor in the economy. This shift will necessitate a fundamental reevaluation of our economic structures, social systems, and the meaning of work in human society. This will also impact the economic value of education.

## 5. Education and obsolescence of human capital

The relationship between technology and education has long been a central focus of economics. Education allows workers to improve their skills and earn a premium in the labor market. Tinbergen (1974) famously described the evolution of this skill premium as the outcome of a "race between education and technology." He observed that the education system determined the supply of skilled labor, while the level of technology determined how much skilled workers are in demand. During the 20$^{th}$ century, he posited that the increasing skill requirements stemming from technological advances outpaced the education system's ability to produce skilled workers, leading to a rising skill premium and widening inequality between skilled and unskilled workers.

Building on this framework, Katz and Murphy (1992) provided empirical evidence that skill-biased technological change (SBTC) significantly drove wage inequality in the United States. While the supply of educated workers increased, it did not keep up with the rapid rise in demand, leading to rising returns to education and higher wage disparities in the second half of the 20$^{th}$ century. Their findings justified greater investment in education and training, which shaped people's mindset towards education and educational policies worldwide.  Promoting investment in higher education was seen as a recipe to boost growth and reduce inequality.

**Obsolescence of skills:** AGI will, by definition, be capable of performing any cognitive task at or above human level, including tasks currently associated with higher education and specialized training, such as critical thinking, complex problem-solving, and even creativity. This would make many of the skills valued by the market in recent decades largely obsolete and would fundamentally undermine the value of traditional education.

Early evidence suggests that even generative AI at the level of GPT-3/GPT-4 already benefits lesser-skilled and lesser-experienced workers more than their highly educated and experienced peers (Brynjolfsson et al., 2023; Noy and Zhang, 2023; Autor, 2024). Put differently, current AI systems level the playing field by making the professional skills and experience of many white-collar professionals less valuable. I view this as the beginning of the skill devaluation that we will experience under AGI.

**Reassessing education:** AGI will thus require a fundamental reorientation of our educational goals and curricula. In the near term, one of the most critical skills will be AI literacy, the ability to collaborate effectively with AI. This involves understanding both AI systems' capabilities and limitations. Education would be most valuable if it focuses on teaching students how to effectively interact with AI tools, interpret their outputs, and make informed decisions based on AI-generated information. In other words, before AGI is realized, understanding when human judgment is necessary in addition to AI assistance, and knowing how to integrate AI into human decision making would be very important skills.

Education must also address the broader implications of AI on society. Students should be equipped to critically evaluate the ethical implications of AI deployment and participate in informed discussions about AI governance. However, given how rapidly AI advances, these skills would need to be constantly upgraded.

Under AGI, the economic value of education will come primarily from facilitating the types of jobs described previously, where human labor is needed for temporary technical and social reasons or for fundamental human-centric reasons. Education for these roles will need to focus on developing adaptable skills for managing technological transitions and human-AI interactions in the short term, while cultivating competencies for roles that require human identity, authentic human connection, religious significance, and human oversight in AI alignment.

Despite the erosion of the economic value of education, its civic value is likely to persist. In an AI-dominated world, understanding technology's societal impacts becomes crucial for effective citizenship. Education should teach students to critically evaluate AI-driven changes and make their voices heard by participating in the public discourse about AI policies. It may also be crucial in developing ethical frameworks for AI governance.

**Delivery of education:** At the same time, AI technologies will also revolutionize the delivery of education (Khan, 2024). AI systems have only recently developed the theory of mind necessary to be effective educators (Kosinski, 2023). This will enable them to identify and fill the gaps in students' grasp of the material, acting as personalized tutors. Moreover, recent AI systems have also developed the ability to simulate human connection, making them potentially more effective as educators.

The role of human educators will be profoundly transformed. Traditionally, educators were valued for their subject matter expertise, particularly in higher education. However, this role will become redundant in an AI-dominated educational landscape – AI systems will be the best subject matter experts. The role of educators will diminish. At best, they will be reduced to becoming facilitators of learning experiences, guiding students through AI-assisted learning processes, particularly for younger students.

**Demand for education:** Concurrent with these changes, a significant decline in the demand for traditional educational services and institutions is likely. Basic education serves broader societal functions beyond skill acquisition and may be less affected. However, higher education will experience profound challenges. In recent decades, higher education has been one of the fastest-growing sectors, driven by parents' insatiable demand to invest in their children's human capital that was based on the 20th-century promise that education was the golden ticket to prosperity.

AGI may dramatically reduce demand for higher education as it devalues cognitive skills and undermines the returns to human capital investment. The economic value from the signaling and certification role of higher education – degrees from well-regarded universities signal student's value as a future worker – will decline, although degrees may continue to provide social status benefits. Moreover, educational institutions will continue to serve other roles, including socialization, personal development, and the transmission of human knowledge and culture.

Simultaneously with the decline in demand, institutions of higher education will face increasing competition from AI-driven alternatives capable of delivering personalized education at scale, potentially offering more cost-effective and flexible learning options. Under this competitive pressure, institutions of higher education will need to restructure their offerings, pivoting away from teaching skills that were once economically valuable towards what will remain in demand in an AGI future.

These adaptations will clash with existing business models that involve long-term commitments, such as tenure to subject matter experts in domains that may decline in relevance in an AGI world. As tenured faculty members have pervasive freedoms in determining their research and teaching focus, the economic pressures that their employers face may create conflict between institutional needs and faculty autonomy.

**Role of universities in research:** Universities and other research institutions will have to grapple with the potential of AGI rendering human researchers redundant. Universities that want to maintain their research relevance will need to make vast investments in technological infrastructure, including computing power, robotics, and other equipment required to conduct research in an AI-dominated landscape. They will also have to either team up with or fend off challenges from AI-powered newcomers in the research space. From a broader perspective, this shift requires universities to substitute cognitive labor with capital, mirroring the trends in other white-collar sectors of the economy.

These changes require a fundamental rethinking of the role and structure of universities. The transition will be challenging and contentious, as it requires balancing adaptation with the preservation of academic values and honoring long-standing commitments to faculty and educational philosophies.

# IV. Challenges for economic policy

Let us now synthesize our findings into eight specific challenges that AGI will pose for economic policy. Addressing these eight challenges will be crucial to ensure that AGI contributes to a human future with shared prosperity.

## Challenge 1: Inequality and income distribution in an AGI world

**AGI will disrupt labor markets and could lead to unprecedented levels of income concentration. Compensating the losers would necessitate new ways of distributing income and economic benefits in society.**

As AGI systems become capable of performing a wide range of cognitive tasks, they could lead to widespread labor displacement, mass unemployment, and stark wage declines (Korinek and Juelfs, 2024). If left to the market, the benefits of AGI would accrue primarily to those who own capital and control AGI technologies. Our current systems of social insurance revolve primarily around work – for example, people receive retirement benefits after having worked for several decades, or they receive unemployment if they lose their jobs. In many countries, health benefits are linked to work. In an AGI future in which labor markets are disrupted, compensating the losers would require new mechanisms for income distribution that are independent of work, such as, for example, Universal Basic Income (UBI). Ultimately, the challenge lies in harnessing the immense potential of AGI to create a more prosperous society for all, rather than allowing it to widen economic divides. However, implementing such systems would require significant changes to current economic models.

The challenge of income inequality in the face of AI-driven technological change becomes even more daunting when we consider between-country rather than within-country disparities. As Korinek and Stiglitz (2021) argue, the impact of AI and related technologies on developing countries could be particularly severe. These countries often rely on their comparative advantage in labor-intensive industries, which may be devalued by AI-driven automation. Unlike within-country redistribution, where national governments have mechanisms to compensate losers, there are no well-established global institutions for large-scale redistribution across countries. The potential for AI to exacerbate global inequality is therefore much greater, and the policy challenges far more complex. Addressing this issue would require unprecedented levels of international cooperation and potentially the development of new global economic governance structures to share the benefits of AI more equitably across nations.

## Challenge 2: Education and skill development

**AGI will render many traditional skills obsolete, requiring a fundamental reevaluation of our educational goals and curricula to prepare individuals for a radically different world.**

Many skills that are currently valued in the job market may become obsolete with the advent of AGI, challenging the role of education as a means to acquire marketable skills and secure employment. In an AGI world, the remaining human jobs may fall into two categories: those preserved by temporary technical and social barriers, and those maintained due to fundamental human-centric aspects (like authenticity of human connection, religious beliefs, or AI oversight). The economic value of human education will likely shift towards facilitating the latter category of jobs, where human interaction is preferred for cultural, emotional, or ethical reasons. Despite the potential erosion of the economic value of education, its civic value is likely to persist. In an AI-dominated world, understanding the technology's societal impacts becomes crucial for effective citizenship. This requires students to learn how to critically evaluate AI-driven changes in society and participate in the public discourse about AI policies. Simultaneously, AI technologies will revolutionize the delivery of education itself. The key challenge lies in transforming education to equip humans for life and coexistence with superintelligent machines.

## Challenge 3: Social and political stability

**AGI-induced labor market disruption could lead to widespread economic discontent, social unrest, and political instability, potentially undermining democratic institutions.**

If AGI leads to widespread labor displacement and stark wage declines, it may create economic insecurity for large segments of the population while also depriving workers of an important element of purpose. This disruption could have profound implications for political stability (Bell and Korinek, 2023). Historically, significant economic displacements have often led to social unrest and the rise of populist or authoritarian movements. In an AGI world, the concentration of economic power in the hands of those who control AGI technologies could exacerbate inequalities, potentially undermining democratic processes and institutions. Moreover, societies may struggle to find new ways to effectively ensure economic security and social inclusion for all. This could lead to increased polarization and conflict between those benefiting from AGI and those left behind. Additionally, the rapid pace of change brought about by AGI could outstrip the ability of social and political institutions to adapt, eroding public trust in democratic systems. Addressing these challenges would require policies to mitigate economic disruptions, ensure equitable distribution of AGI benefits, and strengthen democratic institutions to withstand the pressures of rapid technological change.

## Challenge 4: Macroeconomic considerations

**The advent of AGI necessitates a fundamental rethinking of macroeconomic policy frameworks, including in the realms of aggregate demand management and fiscal policy.**

Traditional macroeconomic frameworks were built around human labor as the primary factor of production – they will require significant adaptation to remain effective in an AGI-driven economy. Aggregate demand in the AGI economy may shift from human consumption to AI-driven investment demand (Korinek, 2019). As AGI systems become more prevalent, demand for inputs like energy, computing resources, and specialized hardware will rise. However, for AGI to benefit humanity broadly, policymakers must implement mechanisms to provide humans with sufficient income, especially if labor markets weaken. This might involve new forms of redistribution, such as universal basic income, and/or novel ways of linking human welfare to AGI productivity gains.

Monetary policy will also face novel challenges as the nature of inflation and economic slack change. The Phillips curve, which has long guided central bank decision-making by relating unemployment to inflation, will need to be fundamentally reconceptualized. Traditional measures of labor market slack may become irrelevant for inflation dynamics. Instead, new indicators of economic capacity utilization, such as the intensity of capital use (including computing resources and robots), may emerge. New economic models will need to be developed, based on different types of data, and with a potentially redefined monetary policy objective.

Fiscal policy will also need to be adapted. Sources of government revenue will have to shift from labor to other bases, such as capital that includes AI-related assets, or to new forms of value-added taxes that capture AI-generated economic activity. This will have to occur at the same time as the social spending needs discussed under Challenge 1 impose growing burdens on budgets. Governments may also have to make significant investments in research and development to ensure that AGI development aligns with human values and societal goals.

All along, policymakers will need to grapple with the potential for rapid and unpredictable technological changes brought about by AGI. This may require more flexible and adaptive policy frameworks that can quickly respond to sudden shifts in productivity, employment patterns, or economic structures. It may also necessitate closer collaboration between economic policymakers and AGI developers to anticipate and mitigate potential macroeconomic disruptions. The key challenge will lie in reimagining macroeconomic policy for an era where AGI, rather than human labor, becomes the primary driver of economic growth and fluctuations.

## Challenge 5: Antitrust and market regulation

**AGI could lead to unprecedented market concentration and power consolidation, necessitating a fundamental rethinking of antitrust policies and market regulation.**

As AGI systems become increasingly capable and economically valuable, there is a risk of extreme market concentration due to the technology's unique characteristics. AGI development requires massive computational resources, vast amounts of data, and highly specialized talent, all of which create substantial barriers to entry and economies of scale (Korinek and Vipra, 2024; Schmid et al., 2024). This could lead to a winner-take-all scenario where a single or a small group of companies dominate the AGI market, potentially extending their power across most sectors of the economy. The challenge for competition authorities is to balance the need for innovation and efficiency with fair competition and lack of monopolistic practices. Traditional antitrust frameworks may prove inadequate in addressing the unique dynamics of AGI markets, such as the potential for rapid market tipping or the implications of vertical integration between AGI developers and large technology companies. Policymakers must grapple with questions of how to promote a competitive ecosystem of providers of goods and services, ensure sufficiently broad access to critical resources like compute and data, and prevent the abuse of market power. This may require new regulatory approaches, such as mandating data sharing, promoting interoperability, or implementing novel forms of oversight for AGI systems. Additionally, regulators must address potential safety risks associated with powerful AI systems. The key challenge lies in developing new antitrust strategies to ensure fair competition while preventing the concentration of technological power.

## Challenge 6: Intellectual property

**AGI will pose significant challenges to existing intellectual property frameworks and raises questions about how to best share the economic benefits of AI-generated innovations.**

The rise of AGI systems capable of engaging in innovation and of generating vast amounts of content challenges our traditional system of intellectual property (IP) rights. Key issues include determining ownership of AI-generated works, redefining concepts of originality and novelty, and reassessing the appropriate duration of IP protections in a manner that both provides sufficient economic incentives and distribution of the benefits of AGI. The economic incentives provided by current IP frameworks may be ill-suited to a world where AGI can rapidly produce innovations. This mismatch could lead to an excessive concentration of economic power among AGI developers and owners.

## Challenge 7: Environmental implications

**The proliferation of AGI may pose significant environmental challenges, necessitating a careful balance between technological advancement and ecological sustainability.**

The energy demands of increasingly powerful AI systems may exacerbate climate change and resource depletion. The computing infrastructure required to train and run advanced AI models consumes substantial amounts of electricity and risks increasing our emissions of carbon and other pollutants. The AGI scenarios outlined in the previous section involve substantial take-offs in economic growth due to wide-spread AI and robot usage and are likely to also involve substantial take-offs in energy usage that may go far beyond the single-digit annual growth rates that are currently projected for advanced countries, risking potentially catastrophic increases in emissions.

Economists have long studied the externalities of different forms of energy production (see, e.g., Nordhaus, 2019). Since the social cost of carbon and other pollutants is convex, vast increases in energy usage may come hand in hand with vast growth in the externalities generated by unclean forms of energy usage. The consequences of not internalizing the associated externalities could transform earth into a climate hellscape.

The silver lining is that AI may also offer solutions to environmental challenges. AGI could accelerate the development of cleaner energy technologies such as nuclear fusion, the design of more sustainable products and processes, and enhance environmental monitoring and conservation efforts. AGI could also revolutionize climate modeling, enabling more accurate predictions of climate change impacts and more effective mitigation strategies. As AGI systems become more autonomous, it will be crucial to align their objectives with long-term environmental sustainability to prevent unintended ecological catastrophes and ensure that planet earth remains hospitable to its current inhabitants. The key challenge will be to balance the environmental costs of AGI development with its potential benefits.

## Challenge 8: Global AI governance

**AGI will present unprecedented challenges to global governance, requiring new international frameworks to manage power dynamics, ensure equitable development, and mitigate existential risks.**

As AGI emerges as a transformative force, it may reshape global power structures and exacerbate existing inequalities between nations. The concentration of AGI capabilities in a few technologically advanced countries and powerful corporations could lead to a new form of a global "intelligence divide," disrupting the current balance of power and potentially destabilizing international relations. This could occur rapidly, outpacing the ability of existing international institutions to effectively regulate and govern AGI's use. Addressing these challenges will require new forms of global cooperation and governance. This includes establishing international AGI development standards, creating specialized multilateral institutions, developing fair distribution mechanisms for AGI benefits, implementing robust verification systems, and coordinating international efforts on AGI safety research (Ho et al., 2023).

A crucial aspect of global governance will be managing AI safety. As AGI systems become more powerful, ensuring their alignment with human values becomes increasingly important and potentially costly. However, whereas the benefits of safe AGI would accrue to all of humanity, the costs would be borne by a few actors (see, e.g., Korinek and Balwit, 2024). This creates a public goods problem that requires international cooperation to solve, necessitating mechanisms to incentivize and fund global AGI safety efforts. Success will require innovative diplomatic efforts, novel governance structures, and a shared recognition of AGI as both a global opportunity and a global risk that transcends national interests. Fostering the required level of international cooperation and trust in the face of a technology that could radically alter the foundations of global power will be a key challenge.

# V. Conclusions and Future Outlook

As we stand at the dawn of the Age of AI, it is crucial to reflect on the profound economic shifts that lie ahead and to proactively prepare economic frameworks that can inform economic policymakers. This paper has explored a range of challenges that could arise from the advent of advanced AI systems, particularly the potential development of AGI.

The transition to an AI-driven economy is likely to bring fundamental changes in our economic paradigm. I anticipate significant shifts in what will be the important factors of production in the economy, with AI and robots potentially becoming perfect substitutes for human labor across a wide range of cognitive and physical tasks. This transformation may lead to unprecedented productivity gains and economic growth, but it also raises concerns about income distribution, labor market disruptions, skill devaluation, and the need for new economic frameworks.

The rapid pace of AI development will necessitate swift and thoughtful adaptation by society. Policymakers may face a wide array of interconnected challenges. These may span from rising inequality and adapting education systems to rethinking effective macroeconomic policy.

## 1. The Role of Economists in Shaping the Future of AI

Economists have a crucial role in preparing for the Age of AI. While our contribution is just one piece of the broader challenge, it is an essential one. Our analysis can provide valuable insights into the economic implications of AI and help guide policy decisions. Moreover, economists can contribute significantly to AI governance and safety efforts. By applying economic principles to the AI alignment problem, we can help develop incentive structures that encourage the development of safe and beneficial AI systems. Furthermore, our understanding of market dynamics and regulatory frameworks can inform the design of effective governance structures for AI.

As we grapple with the described challenges, AI also has the potential to be a powerful tool to support our efforts and make us more productive in research (Korinek, 2023a; 2024). Advanced AI systems can enhance our ability to tackle complex problems. By processing vast amounts of data and identifying complex patterns better than humans, AI systems will eventually be able to help us develop more accurate and nuanced economic models. This enhanced predictive power may also help policymakers devise more effective and targeted interventions.

## 2. Looking to the Future

As we look to the future, we must examine what the potential economic and social implications will be and how to reconcile the rise of artificial intelligence with continued human flourishing. We must anticipate and prepare for a range of possibilities, from scenarios where narrow AI remains a powerful but limited tool to those where AGI becomes the dominant driver of economic activity (Korinek, 2023b). Different trajectories of AI development will have different impacts on our economic systems and social structures. Each path will require different policy approaches and societal adaptations. As AI systems become more sophisticated, we may also need to grapple with questions about the moral status and rights of AI. This philosophical challenge has significant economic implications, potentially affecting how we value AI labor, allocate resources, and structure our economic institutions (Bostrom, 2024).

The challenges and opportunities presented by the Age of AI demand urgent efforts to develop appropriate economic research and policy frameworks. We have limited time to prepare, and we need to collaborate across disciplines – with AI researchers, social scientists, and philosophers – to develop comprehensive approaches to AI governance and economic policy.

In conclusion, the advent of advanced AI presents both unprecedented challenges and extraordinary opportunities for our economy and society. By proactively addressing the policy challenges outlined in this paper – and leveraging AI as a tool for economic problem-solving – we can work towards a future where the benefits of AI are broadly shared, and humans can flourish. The path ahead is uncertain, but with careful analysis, thoughtful policymaking, and a commitment to ethical principles, we can shape an AI-driven future that aligns with economic prosperity and societal values.